\documentclass[12pt]{iopart}

\usepackage[T1]{fontenc}
\usepackage{color}
 \expandafter\let\csname equation*\endcsname\relax
 \expandafter\let\csname endequation*\endcsname\relax
\usepackage{amsmath}
\usepackage{bbm}

\input epsf
\usepackage{iopams}  

\usepackage[pdftex]{graphicx}

\begin{document}

\title[Incorporation of charge- and pair-density-wave states into the one-band model...]{Incorporation of charge- and pair-density-wave states into the one-band model of d-wave superconductivity}

\author{M Zegrodnik$^{1*}$ and J Spa\l ek$^{2\dagger}$}

\address{$^2$Academic Centre for Materials and Nanotechnology, AGH University of Science and Technology, Al. Mickiewicza 30, 30-059 Krak\'ow,
Poland\\
$^2$Marian Smoluchowski Institute of Physics, Jagiellonian University, {\L}ojasiewicza  11, 30-348 Krak\'ow, Poland}

\ead{*michal.zegrodnik@agh.edu.pl, $\dagger$jozef.spalek@uj.edu.pl}
\begin{abstract}
We study the coexistence of pair- (PDW) and charge-density-wave (CDW) states within the single-band $t$-$J$-$U$ and Hubbard models of $d$-$wave$ superconductivity and discuss our results in the context of the experimental observations for the copper-based compounds. In order to take into account the correlation effects with a proper precision, we use the approach based on the \textit{diagrammatic expansion of the Gutzwiller wave function} (DE-GWF), that goes beyond the renormalized mean field theory (RMFT) in a systematic manner. According to our analysis of the $t$-$J$-$U$ model, the transition between the pure $d$-$wave$ superconducting phase (SC) and the coexistent CDW+PDW phase takes place at $\delta\approx 0.18$ (close to the optimal doping), with the modulated phase located in the underdoped regime. The situation is slightly different for the case of the Hubbard model, where a narrow stability regime of a precursor nematic phase sets in preceding the formation of the modulated CDW+PDW state, with the decreasing hole doping. The results complete our discussion of the standard phase diagram for high-T$_C$ superconducting compounds within the DE-GWF variational approach in the single narrow-band case.
\end{abstract}


\maketitle

\section{Introduction}\label{sec:intro}
Charge-density-wave (CDW) state plays an important role in the physics of underdoped copper-based high-T$_C$ superconducting (SC) compounds. For the ytrium- (YBCO), as well as Bi- and Hg- based materials, the icommensurate CDW modulation vectors $\mathbf{Q}$ lie in the copper-oxide plane and have the form $(0,Q)2\pi/a$ and $(Q,0)2\pi/a$ (with $a$ being the Cu-O lattice constant) \cite{Wu2015-Nature,Achkar2012-PRL,Ghiringhelli2012-Science,Chang2012,Tabis2014}, with a weakly doping dependent periodicity $Q\sim0.31$ for YBCO \cite{Achkar2012-PRL,Ghiringhelli2012-Science, Tabis2014,Blackburn2013-PRL} and $Q\sim 0.26$  ($0.3$) for the single-layer (double-layered) Bi-based compounds \cite{Comin2014-Science,daSilva2014-Science}, as well as $Q\sim 0.27$ for Hg1201 \cite{Tabis2014}. It is still under debate whether the proximity of the vector $\mathbf{Q}$ to the vectors connecting neighboring hot spots at the Fermi surface is only a coincidence, or in fact, the Fermi surface topology imposes the CDW periodicity. Both the charge-stripes with a $90^o$ rotated domains and the checkerboard pattern are consistent with the two simultaneously measured modulation vectors and it is not settled as yet, which scenario is realized experimentally. Nevertheless, some reports point to the charge-stripes as the actual form of the CDW state \cite{Comin2015}.

For the La-based cuprates, where the charge order has been initially observed, it has been confirmed that the uniaxial stripe domains with periodicity of $\sim 4a$ are forme \cite{Tranquada1995,Fink2011,Hucker2013}. However, in this case an arrangement of combined charge and spin order is believed to appear simultaneously. Another difference between the La-based materials and YBCO is that for the latter the modulation vector increases slightly with the decreasing doping, whereas for the former the opposite is true \cite{Blanco2014}. Also, in Bi-, Y-, and Cl-based cuprates it is believed that the $d$-$wave$ bond-order component of the modulation is the dominant one \cite{Comin2015}, whereas for the La-based materials the experimental study reveal the predominant $s'$-$wave$ bond-order \cite{Achkar2016}. Note that the $d$-$wave$ bond order preserves the nodality of the diagonal direction in the Cu-O plane. As one can see, there are some significant differences between the majority of the cuprate family and the La-based cuprates when it comes to the features of the charge ordered phase. In our considerations we refer mainly to the former.  
 

Since both the high-temperature superconductivity and CDW phase appear in the same doping range of the phase diagram (the underdoped regime), the natural question concerns the interplay between the two phases. The experimental data clearly show the competition between the CDW and SC what is manifested by the plateau in the SC critical temperature in the underdoped regime, as well as by the suppression of the CDW intensity peak and correlation length below T$_C$ \cite{Chang2012,Tabis2014,Hucker2014,Wu2011,Ghiringhelli2012}. Another issue which concerns the relationship between the two phases is the possibility of spatially modulated Cooper-pair density, which could coexist with charge ordering in the underdoped regime. Such pair-density-wave state (PDW) has some principal similarities with the Fulde-Ferrell-Larkin-Ovchinnikov state, which has been proposed in various systems \cite{Croitoru2012, Wu2013, Croitoru2014, Ptok2014, Wojcik2015}. Very recently, the PDW state has been reported experimentally in BSCCO \cite{Hamidian2016}. Moreover, it has been found that the pair-density and charge-density modulations are governed by very similar vectors $\mathbf{Q}\approx(0.25,0)2\pi/a$ and $\mathbf{Q}\approx(0,0.25)2\pi/a$.



Here, we analyze theoretically the PDW and CDW states coexistence within the single-band Hubbard and $t$-$J$-$U$ models with the inclusion of correlation effects by going beyond the RMFT approach in a systematic manner. Namely, we use the \textit{diagrammatic expansion of the Gutzwiller wave function} (DE-GWF) method, which allows us to obtain the full Gutzwiller wave function solution for the modulated states and study their stability as a function of doping. It has been shown previously that by accounting for the electron correlations already at the level of RMFT one can obtain the proper charge ordering modulations with a dominant $d$-form factor within the $t$-$J$-$V$ model \cite{Allais2014}. In these considerations the intersite Coulomb interaction term ($\sim V$) is neccesary to induce the charge ordered phase. However, a similar result can also be obtained by using alternative approaches \cite{Chowdhury2014,Wang2014,Meier2014,Wang2015_CDW}. 

As we show below, by using the DE-GWF approach, the CDW phase appears already within the Hubbard model (with intrasite repulsion only), which is one of the canonical one-band models used for the description of the Cu-O planes in the copper-based materials. Furthermore, we study also the $t$-$J$-$U$ model, for which we have recently obtained very good quantitative agreement between theory and experiment for the selected principal observations of the $d$-$wave$ superconducting state in the cuprates. With this analysis, we extend our previous considerations of pure $d$-$wave$ superconductivity to the description of both charge- and Cooper-pair-modulated states. 
Within our study both the PDW and CDW states are modulated according to a fixed single commensurate vector $\mathbf{Q}=(1/3,0)2\pi/a$, which is close to the incommensurate one measured in experiment \cite{Chang2012,Comin2014-Science,Comin2015}. Such approach is justified by the very weak measured doping dependence of $\mathbf{Q}$ \cite{Tabis2014} and close proximity of the PDW and the CDW modulation vectors reported in experiments \cite{Hamidian2016}. In our calculations we allow for both the site-centered ($s$-$wave$) and bond-centered ($d$-$wave$, $extended$ $s$-$wave$) CDW. The former corresponds to a situation in which the on-site electron concentration is modulated, whereas for the latter the electron hopping value is modulated. 

It should be noted that for the case of the Hubbard model, the variational cluster approximation has been used recently \cite{Faye2017} to study the stability of the charge-density-wave and pair-density wave states (as well as their coexistence) as a function of hole doping. However, the proper sequence of phases appearing in the experimental phase diagram is different. The PDW+CDW coexistent phase has also been analyzed in terms of the spin-fermion model \cite{Wang2015_PDW}. Within this approach it is argued that the PDW+CDW phase should appear in the underdoped regime, in accordance to the experimental findings.

We show that for the model with both small but non-zero number of double occupancies and the intersite exchange interaction term included explicitly ($t$-$J$-$U$ model), the stability of the coexistent PDW+CDW modulated state is contained in the underdoped regime and the pure $d$-$wave$ SC phase occurs at and above the optimal doping in the phase diagram, which reproduces the experimental situation. However, the issue of modulation form-factor still seems to remain problematic since in our analysis the site-centered CDW contribution appears to be significant, in contradiction to the experimental data, where the nodal, $d$-$wave$ type, modulation persists to the lowest doping possible in the metallic phase.

The structure of the paper is as follows. In the next section we present theoretical model and its solution. Sec. 3 contains detailed numerical results and their discussion. The conclusions are contained in the last Section.

\section{Theory}
\subsection{Model and wave function}
The Hamiltonian considered here is given below
\begin{equation}
\begin{split}
\mathcal{\hat{H}}&=t\sum_{\langle ij\rangle\sigma}\hat{c}^{\dagger}_{i\sigma}\hat{c}_{j\sigma}
+t'\sum_{\langle\langle ij\rangle\rangle\sigma}\hat{c}^{\dagger}_{i\sigma}\hat{c}_{j\sigma}\\
&+J\sum_{\langle ij\rangle}\hat{\mathbf{S}}_i\cdot\hat{\mathbf{S}}_j
+U\sum_i \hat{n}_{i\uparrow}\hat{n}_{i\downarrow},
 \label{eq:H_start}
 \end{split}
\end{equation}
where the first two terms correspond to the single electron hoppings, the third represents the antiferromagnetic superexchange interaction, and the last refers to the intrasite Coulomb repulsion. By $\langle...\rangle$ and $\langle\langle...\rangle\rangle$ we denote the summations over the nearest-neighbors and next nearest-neighbors, respectively. For $J\equiv0$ we obtain the Hubbard model which is also considered here, whereas for the case of $U\rightarrow \infty$ (i.e., $U\gg |t|$) one reproduces the $t$-$J$ model. Formally, this model describes an interpolation between the Hubbard and $t$-$J$ model limits. Physically, it extends the concept of kinetic exchange to the situation when $U$ is not too large as compared to the bare bandwidth $W$. 

This extended model has been discussed in detail by us earlier, as well applied to a quantitative analysis of selected universal properties of cuprate high-T$_C$ superconductors within our original diagrammatic solution of the Gutzwiller wave function in two dimensions \cite{Spalek2017,Zegrodnik2017_1,Zegrodnik2017_2}. Also, the effect of nematicity on the resulting phase diagram has been discussed by us recently \cite{Zegrodnik2017_nematic}. In what follows we supplement our extensive analysis with the incorporation of CDW/PDW solution into the scheme.

In order to take into account inter-electronic correlations we use the DE-GWF approach to the Gutzwiller-type wave function defined by
\begin{equation}
 |\Psi_G\rangle\equiv\hat{P}_G|\Psi_0\rangle,
 \label{eq:Gutz_wave_func}
\end{equation}
where $|\Psi_0\rangle$ is the non-correlated wave function (to be defined later) and the correlation operator $\hat{P}_G$ is provided below 
 \begin{equation}
 \hat{P}_G\equiv\prod_i\hat{P}_i=\prod_i\sum_{\Gamma}\lambda_{i,\Gamma} |\Gamma\rangle_{ii}\langle\Gamma|,
  \label{eq:Gutz_operator}
\end{equation}
where $\lambda_{i,\Gamma}\in\{\lambda_{i\emptyset},\lambda_{i\uparrow},\lambda_{i\downarrow},\lambda_{i d}\}$ are the variational parameters which correspond to four states from the local basis $|\emptyset\rangle_i\;, |\uparrow\rangle_i\;, |\downarrow\rangle_i\;, |\uparrow\downarrow\rangle_i$ at site $i$, respectively.

An important step of the DE-GWF method is the application of the condition \cite{Gebhard1990,Bunemann2012} 
\begin{equation}
 \hat{P}_i^2\equiv1+x_i\hat{d}^{\textrm{HF}}_i,
 \label{eq:constraint}
 \end{equation}
 where $x_i$ is yet another variational parameter and $\hat{d}^{\textrm{HF}}_i\equiv\hat{n}_{i\uparrow}^{\textrm{HF}}\hat{n}_{i\downarrow}^{\textrm{HF}}$, $\hat{n}_{i\sigma}^{\textrm{HF}}\equiv\hat{n}_{i\sigma}-n^{(0)}_{i\sigma}$, with $n^{(0)}_{i\sigma}\equiv\langle\Psi_0|\hat{n}_{i\sigma}|\Psi_0\rangle$. One should note that $\lambda_{i,\Gamma}$ parameters for a given site $i$ are functions of $x_i$ which means that there is a single variational parameter per atomic site in such an approach. As it has been shown in Refs. \cite{Gebhard1990,Bunemann2012}, condition (\ref{eq:constraint}) leads to a rapid convergence of the resulting diagrammatic expansion with the increasing order of the variational parameter $x_i$. For the case of a spatially homogeneous state one has $x_i\equiv x$. The formulation and discussion of the DE-GWF approach for the case of homogeneous $d$-$wave$ superconducting or paramagnetic states is provided in Refs. \cite{Gebhard1990,Bunemann2012,Kaczmarczyk2014,Kaczmarczyk2016}. For the case of the CDW/PDW states the $x_i$ parameter follows the modulation which characterizes those new ordered phases. In what follows we denote the normalized expectation values in the correlated state by $\langle...\rangle_G=\langle\Psi_G|...|\Psi_G \rangle/\langle\Psi_G|\Psi_G\rangle$, while the corresponding expectation values in the non-correlated state $\langle...\rangle_0=\langle\Psi_0|...|\Psi_0\rangle$. 

\subsection{CDW and PDW states: A general characterization}
In order to encompass both site- and bond-centered charge orderings, as well as the PDW phase appearance, one has to allow for a modulation of the hopping averages $\langle\hat{c}^{\dagger}_{i\sigma}\hat{c}_{j\sigma}\rangle_G$, electron concentration $\langle\hat{n}_i\rangle_G$, and pairing averages $\langle\hat{c}^{\dagger}_{i\uparrow}\hat{c}^{\dagger}_{j\downarrow}\rangle_G$ in the considered wave function. Assuming that all three types of averages are modulated by a single vector one can write that
\begin{equation}
\langle\hat{c}^{\dagger}_{i\sigma}\hat{c}_{j\sigma}\rangle_G=\bar{P}_{ \mathbf{g}_{ij}}+\delta P_{\mathbf{g}_{ij}}\cos \big[\mathbf{Q}(\mathbf{R}_{j}+\mathbf{g}_{ij}/2)\big],
\label{eq:mod_P}
\end{equation}
\begin{equation}
\langle\hat{n}_{i\sigma}\rangle_G=\bar{n}+\delta n_{CDW}\cos \big[\mathbf{Q}\mathbf{R}_{i}\big],
\label{eq:mod_n}
\end{equation}
\begin{equation}
\langle\hat{c}^{\dagger}_{i\uparrow}\hat{c}^{\dagger}_{j\downarrow}\rangle_G=\bar{\Delta}_{\mathbf{g}_{ij}}+\delta \Delta_{\mathbf{g}_{ij}}\cos \big[\mathbf{Q}(\mathbf{R}_{j}+\mathbf{g}_{ij}/2)\big],
\label{eq:mod_S}
\end{equation}
where $\mathbf{g}_{ij}=\mathbf{R}_i-\mathbf{R}_j$ and $\mathbf{Q}$ is the modulation vector. Note that the reference values $\bar{P}_{\mathbf{g}_{ij}}$,  $\bar{\Delta}_{\mathbf{g}_{ij}}$ and the modulation amplitudes $\delta P_{\mathbf{g}_{ij}}$, $\delta \Delta_{\mathbf{g}_{ij}}$ depend only on the vector connecting sites $i$ and $j$. Eq. (\ref{eq:mod_P}) and (\ref{eq:mod_S}) corresponds to $i\neq j$ only since we do not include the intrasite pairing in our approach. As one can realize, in that situation the solution will contain a number of self-consistent integral equations and thus the computations are quite involved.

\begin{figure}[h!]
\centering
\includegraphics[width=1.0\textwidth]{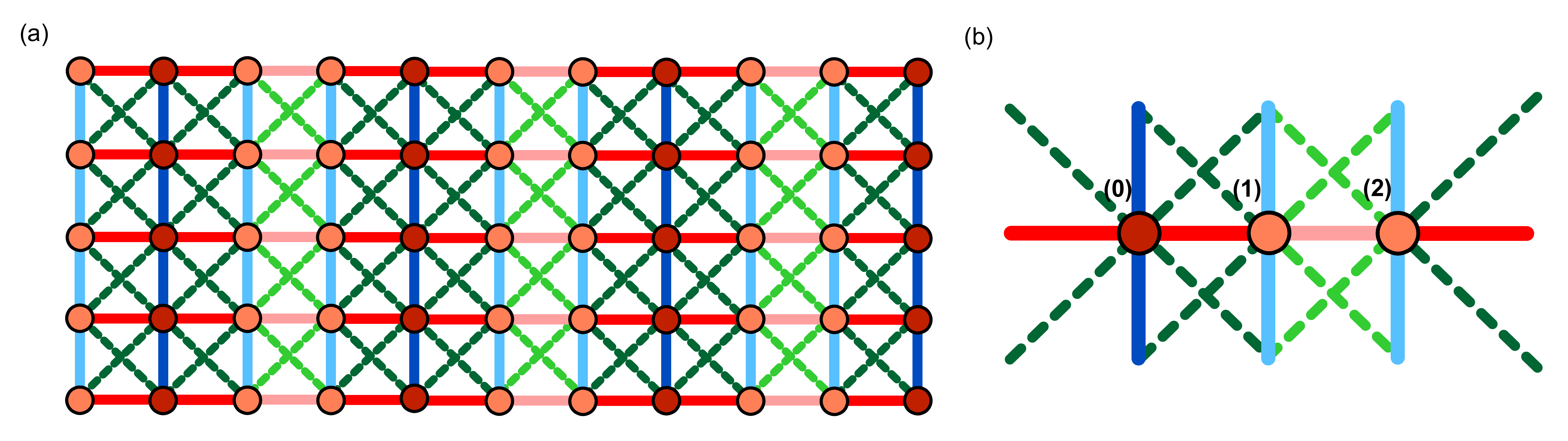}
\caption{(a) Schematic illustration of charge modulation on square lattice with the vector $\mathbf{Q}=(1/3,0)2\pi/a$. Dots and lines with different color correspond to different values of $\langle\hat{n}_{i\sigma}\rangle$ and $\langle\hat{c}^{\dagger}_{i\sigma}\hat{c}_{j\sigma}\rangle$ averages, respectively. The nearest-neighbor hopping averages are marked by the solid lines while the next-nearest neighbor averages are marked by the dashed lines. In (b) we show the three atomic-sites which compose the repeating pattern and are enumerated by (0), (1), and (2). Concentrations on sites (1) and (2) are equal and different from that on site (0).}
\label{fig:lattice}
\end{figure}

As already stated in the preceding section, here we represent the modulation by a single commensurate vector in the form $\mathbf{Q}=(1/3,0)2\pi/a$, which is close to the incommensurate one measured in experiments \cite{Chang2012,Comin2014-Science,Comin2015}. Such a choice leads to a modulation along the $x$ axis with the period of $3a$ in real space. Schematic illustration of how the electron concentration and hopping averages change in real space is provided in Fig. \ref{fig:lattice}. As one can see, there is a repeating pattern of three consecutive atomic sites labeled by (0), (1), and (2) [cf. Fig. \ref{fig:lattice} (b)]. Atomic sites (1) and (2) have the same value of $\langle\hat{n}_{i\sigma}\rangle$, which differs from that on site (0). Analogous repeating pattern can be found for the hopping averages (marked by the solid and dashed lines) and the pairing averages (not shown in the Figure for the sake of clarity). 

In our diagrammatic approach we assume that the hopping and pairing averages in the non-correlated ($|\Psi_0\rangle$) and the correlated ($|\Psi_G\rangle$) states can have non-zero values up to the fourth nearest neighbor. This represents the real-space cutoff which is going to be discussed briefly in the next subsection when the diagrammatic method is described. All the non-zero averages taken into account are modulated according to Eqs. (\ref{eq:mod_P})-(\ref{eq:mod_S}). However, the dominant contribution to the wave function comes from the nearest and next-nearest neighbor contributions. Therefore, for the sake of clarity we focus here on the analysis of the modulations of the on-site electron concentration $\langle\hat{n}_{i\sigma}\rangle_G$, the nearest and next-nearest neighbor hopping averages $\langle\hat{c}^{\dagger}_{i\sigma}\hat{c}_{j\sigma}\rangle_G$, as well as the nearest neighbor pairing averages $\langle\hat{c}^{\dagger}_{i\uparrow}\hat{c}^{\dagger}_{j\downarrow}\rangle_G$ (since the diagonal next-nearest neighbor pairing contribution is zero due to the $d$-$wave$ symmetry of the pairing).

For the selected modulation vector it is convenient to use the following site-dependent hopping and pairing parameters representing the considered symmetries of the nearest- and next-nearest neighbor averages 
\begin{equation}
P^{d,s',x}_i=\frac{1}{4}\sum_{\langle j(i)\rangle}\gamma_{ij}^{d,s',x}\langle\hat{c}^{\dagger}_{j\sigma}\hat{c}_{i\sigma} \rangle_G,\quad P^{s'',x'}_i=\frac{1}{4}\sum_{\langle\langle j(i)\rangle\rangle}\gamma_{ij}^{s'',x'}\langle\hat{c}^{\dagger}_{j\sigma}\hat{c}_{i\sigma} \rangle_G,
\label{eq:mod_Pdsx_def}
\end{equation}
\begin{equation}
\Delta^{d,s',x}_i=\frac{1}{4}\sum_{\langle j(i)\rangle}\gamma_{ij}^{d,s,x}\langle\hat{c}^{\dagger}_{j\uparrow}\hat{c}^{\dagger}_{i\downarrow} \rangle_G,
\label{eq:mod_Sdsx_def}
\end{equation}
where $\langle j(i) \rangle$ ($\langle\langle j(i)\rangle\rangle$) denotes the summation over the nearest (next-nearest) neighbors of atomic-site $i$. The symmetry factors are defined below
\begin{equation}
\begin{split}
\gamma^d_{ij}&=(\delta_{\mathbf{g}_{ij}-\hat{x}}+\delta_{\mathbf{g}_{ij}+\hat{x}}-\delta_{\mathbf{g}_{ij}-\hat{y}}-\delta_{\mathbf{g}_{ij}+\hat{y}}),\\
\gamma^{s'}_{ij}&=(\delta_{\mathbf{g}_{ij}-\hat{x}}+\delta_{\mathbf{g}_{ij}+\hat{x}}+\delta_{\mathbf{g}_{ij}-\hat{y}}+\delta_{\mathbf{g}_{ij}+\hat{y}}),\\
\gamma^x_{ij}&=(\delta_{\mathbf{g}_{ij}-\hat{x}}-\delta_{\mathbf{g}_{ij}+\hat{x}}),\\
\gamma^{s''}_{ij}&=(\delta_{\mathbf{g}_{ij}-\hat{x}-\hat{y}}+\delta_{\mathbf{g}_{ij}+\hat{x}+\hat{y}}+\delta_{\mathbf{g}_{ij}+\hat{x}-\hat{y}}+\delta_{\mathbf{g}_{ij}+\hat{x}+\hat{y}}),\\
\gamma^{x'}_{ij}&=(\delta_{\mathbf{g}_{ij}-\hat{x}-\hat{y}}-\delta_{\mathbf{g}_{ij}+\hat{x}+\hat{y}}-\delta_{\mathbf{g}_{ij}+\hat{x}-\hat{y}}+\delta_{\mathbf{g}_{ij}-\hat{x}+\hat{y}}),\\
\end{split}
\label{eq:symmetries}
\end{equation}
with $\delta_{\mathbf{v}}$ being the appropriate Kronecker delta. The parameters $P^d_{i}$ ($\Delta^d_{i}$), $P^{s'}_{i}$ ($\Delta^{s'}_{i}$), and $P^{s''}_{i}$ correspond to the $d$-$wave$ and $extended$ $s$-$wave$ hopping (pairing) contributions to the wave function, respectively. Note also that the non-zero values of both $P^d_i$ ($\Delta^d_i$) and $P^{s'}_i$ ($\Delta^{s'}_i$) leads to breaking of the $C_4$ symmetry which, still does not imply the presence of charge ordering, as such condition can be fulfilled also in the homogeneous nematic (coexistent nematic-superconducting) state. The CDW (PDW) phase appears only when $P^{d,s',s''}_{i}$ ($\Delta^{d,s'}_{i}$) becomes site-dependent according to the modulation $\mathbf{Q}$. In such a situation the hopping (pairing) averages from the atomic sites (1) and (2) to the left-hand-side neighbor can be different than the corresponding right-hand-side hopping (pairing), what leads to non-zero values of $P^x_{i}$ ($\Delta^x_{i}$) (cf. Fig. \ref{fig:lattice}). The latter rule also applies to the next-nearest neighbors which results in non-zero values of $P^{x'}_{i}$. The pairing parameters analogical to $P^{s''}_{i}$ and $P^{x'}_{i}$ do not appear, since the pairing in the diagonal direction is zero.

Using Eqs. (\ref{eq:mod_P})-(\ref{eq:mod_Sdsx_def}) one can write that
\begin{equation}
{P}^{d,s',s''}_i=\bar{P}^{d,s',s''}+\delta P^{d,s',s''}_{CDW}\cos \big[\mathbf{Q}\mathbf{R}_{i}\big],
\label{eq:mod_P_dsp}
\end{equation}
\begin{equation}
{\Delta}^{d,s'}_i=\bar{\Delta}^{d,s'}+\delta \Delta^{d,s'}_{PDW}\cos \big[\mathbf{Q}\mathbf{R}_{i}\big],
\label{eq:mod_delt_ds}
\end{equation}
\begin{equation}
{P}^{x,x'}_i=\delta P^{x,x'}_{CDW} \sin \big[\mathbf{Q}\mathbf{R}_{i}\big],
\label{eq:mod_P_p}
\end{equation}
\begin{equation}
\Delta^{x}_i=\delta \Delta^x_{PDW} \sin \big[\mathbf{Q}\mathbf{R}_{i}\big],
\label{eq:mod_delt_p}
\end{equation}
where $\bar{P}^{d,s',s''}$, $\bar{\Delta}^{d,s'}$ are the site-independent reference values and $\delta P^{d,s',s'',x,x'}_{CDW}$, $\delta \Delta^{d,s',x}_{PDW}$ are the symmetry-resolved modulation amplitudes. The amplitudes appearing in Eqs. (\ref{eq:mod_P}) and (\ref{eq:mod_S}) can be expressed by $\delta P^{d,s',s'',x,x'}_{CDW}$ and $\delta \Delta^{d,s',x}_{PDW}$ in the following manner
\begin{equation}
\delta P_{\mathbf{g}_{ij}}=\delta P^d_{CDW}\;\gamma_{ij}^d+\delta P_{CDW}^{s'}\;\gamma_{ij}^{s'}+2\delta P_{CDW}^{x}\;\gamma_{ij}^{x}+\delta P_{CDW}^{s''}\;\gamma_{ij}^{s''}+\delta P_{CDW}^{x'}\;\gamma_{ij}^{x'},
\label{eq:sym_factors_hop}
\end{equation}
\begin{equation}
\delta \Delta_{\mathbf{g}_{ij}}=\delta \Delta^d_{PDW}\;\gamma_{ij}^d+\delta \Delta^{s'}_{PDW}\;\gamma_{ij}^{s'}+2\delta \Delta^x_{PDW}\;\gamma^x_{ij},
\label{eq:sym_factors_pairing}
\end{equation}
from which we can see the resulting modulation can be expressed as a mixture of the considered symmetry contributions. The same applies to the reference values $\bar{P}^{d,s',s'',x,x'}_{CDW}$ and $\bar{\Delta}^{d,s',x}_{PDW}$.

Non-zero value of $\delta P^{d,s',s'',x,x''}_{CDW}$ in Eq. (\ref{eq:sym_factors_hop}) corresponds to the bond-centered CDW, while non-zero value of $\delta n_{CDW}$ in Eq. (\ref{eq:mod_n}) is responsible for the site-centered CDW. The remaining modulation amplitudes $\delta {\Delta}^{d,s',x}_{PDW}$ in Eq. (\ref{eq:sym_factors_pairing}) introduce the PDW phase. All those modulation amplitudes play the role of order-parameter components of the CDW and/or PDW states.

\subsection{Solution methodology}

For the correlated wave function $|\Psi_G\rangle$ with the selected modulation, the expectation value from any two local operators, $\hat{o}_i$ and $\hat{o}^{\prime}_j$ appearing in the initial Hamiltonian (\ref{eq:H_start}), can be expressed in the form (cf. Refs. \cite{Zegrodnik2017_1,Kaczmarczyk2014})
 \begin{equation}
  \langle\Psi_G|\hat{o}_{i}\hat{o}^{\prime}_{j}|\Psi_G\rangle=\sum_{k=0}^{\infty}\frac{1}{k!}\sideset{}{'}\sum_{l_1...l_k}x_0^{k_0}x_1^{k_1}x_2^{k_2}\langle\Psi_0| \tilde{o}_{i}\tilde{o}^{\prime}_{j}\;\hat{d}^{\textrm{HF}}_{l_1...l_k}|\Psi_0 \rangle,
\label{eq:expansion}
\end{equation}
where $\tilde{o}_{i}\equiv\hat{P}_i\hat{o}_{i}\hat{P}_{i}$, $\tilde{o}^{\prime}_{j}\equiv\hat{P}_j\hat{o}^{\prime}_{j}\hat{P}_{j}$, $\hat{d}^{\textrm{HF}}_{l_1...l_k}\equiv\hat{d}^{\textrm{HF}}_{l_1}...\hat{d}^{\textrm{HF}}_{l_k}$, and  $\hat{d}^{\textrm{HF}}_{\varnothing}\equiv 1$. The primed summation has the restrictions: $l_p\neq l_{p'}$, $l_p\neq i,j$ for all $p$ and $p'$. The variational parameters $x_0$, $x_1$, and $x_2$ correspond to the three atomic positions from the repeating pattern depicted in Fig. \ref{fig:lattice}. Since the atomic sites (1) and (2) are equivalent, one can take $x_1\equiv x_2$. For a given term of the summation over $l_1...l_k$, the powers $k_0$, $k_1$, and $k_2$ represent how many times in the set $l_1...l_k$ the indices corresponding to the (0), (1), and (2) appear, respectively. They fulfill the relation $k_0+k_1+k_2=k$. It has been shown \cite{Kaczmarczyk2014,Abram2017} that the desirable convergence can be achieved by taking the first 4-6 terms of the summation over $k$ appearing in Eq.(\ref{eq:expansion}). The results presented in the subsequent section have been obtained by including the terms up to the third order in $k$.

The averages in the non-correlated state $|\Psi\rangle_0$ on the right-hand side of Eq. (\ref{eq:expansion}) can be decomposed with the use of the Wick's theorem and expressed in terms of the correlation functions $P^{(0)}_{ij\sigma} \equiv \langle \hat{c}^{\dagger}_{i\sigma} \hat{c}_{j\sigma}\rangle_0$ and $\Delta^{(0)}_{ij} \equiv \langle \hat{c}^{\dagger}_{i\uparrow} \hat{c}^{\dagger}_{j\downarrow}\rangle_0$. Such a procedure allows us to express the ground state energy $\langle\mathcal{\hat{H}}\rangle_G\equiv\langle\Psi_G|\mathcal{\hat H}|\Psi_G\rangle/\langle\Psi_G|\Psi_G\rangle$ as a function of $P^{(0)}_{ij\sigma}$, $\Delta^{(0)}_{ij}$, $n^{(0)}_{i\sigma}$, and $x_m$. In practice, it is neccesary to introduce the real-space cut-off for the $P^{(0)}_{ij\sigma}$ and $\Delta^{(0)}_{ij}$ parameters. Here, in order to carry out the calculations in a reasonable time, the maximum distance has been taken as $R^2_{\mathrm{max}}=5a^2$, which means that we include the hopping and pairing averages up to the fourth nearest-neighbor. 

Having an explicit formula for the energy expectation values in different phases one can derive the effective Schr\"odinger equation and the set of self-consistent equations for the parameters $P^{(0)}_{ij\sigma}$, $\Delta^{(0)}_{ij}$, in an analogical manner as in Refs. \cite{ Zegrodnik2017_1,Kaczmarczyk2014}. The set of equations is solved in conjunction with the minimization of the energy with respect to $x_0$ and $x_1\equiv x_2$. Having the values of $P^{(0)}_{ij\sigma}$, $\Delta^{(0)}_{ij}$, $x_0$, $x_1$, $x_2$, and $n^{(0)}_{i\sigma}$, one can calculate the correlated pairing averages $\langle\hat{c}^{\dagger}_{i\uparrow}\hat{c}^{\dagger}_{j\downarrow} \rangle_G$ and the correlated hopping averages $\langle\hat{c}^{\dagger}_{i\sigma}\hat{c}_{j\sigma} \rangle_G$. The latter values are, in turn, used in order to obtain the site-independent reference values $\bar{P}^{d,s,s',x,x'}_{CDW}$, $\bar{\Delta}^{d,s,x}_{PDW}$ and the modulation amplitudes $\delta P^{d,s,s',x,x'}_{CDW}$, $\delta \Delta^{d,s,x}_{PDW}$, $\delta n_{CDW}$ in the correlated state $|\Psi\rangle_G$. 

One should note that when it comes to the modulated states we assume that the hopping and pairing averages can be expressed by Eq. (\ref{eq:mod_P}) and (\ref{eq:mod_S}) with $\mathbf{Q}=(1/3,0)2\pi$ and that there is no spontaneous current created. Also, since the PDW phase emerges from the pure $d$-$wave$ paired phase we set the diagonal pairing averages to zero. We calculate all the listed symmetry factors with no further constrictions. Therefore, the relative balance between the particular symmetry contributions results explicitly from the calculations. The following phases appear as stable within the analyzed approach and are discussed in the subsequent Section:
\begin{itemize}
\item Paramagnetic phase (PM): $P^{s'}_i\equiv\bar{P}^{s'}\neq 0$, $P^{s''}_i\equiv\bar{P}^{s''}\neq 0$, $P^{d,x,x'}_i= 0$, $\Delta^{d,s',x}_i= 0$, $\delta P^{d,s',s'',x,x'}_{CDW}=\delta \Delta^{d,s',x}_{PDW}= 0$

\item Pure $d$-$wave$ superconducting phase (SC):  $P^{s'}_i\equiv\bar{P}^{s'}\neq 0$, $P^{s''}_i\equiv\bar{P}^{s''}\neq 0$, $\Delta^d_i\equiv\bar{\Delta}^d\neq 0$, $P^{d,x,x'}_i=0$, $\Delta^{s',x}_i=0$, $\delta P^{d,s',s'',x,x'}_{CDW}=\delta \Delta^{d,s',x}_{PDW}=0$

\item Coexistent superconducting-nematic phase (SC+N): $P^{s'}_i\equiv\bar{P}^{s'}\neq 0$, $P^{s''}_i\equiv\bar{P}^{s''}\neq 0$, $P^d_i\equiv \bar{P}^d\neq 0$, $\Delta^d_i\equiv\bar{\Delta}_d\neq 0$, $\Delta^{s'}_i\equiv \bar{\Delta}^{s'}\neq 0$, $P^{x,x'}_i\equiv 0$, $\Delta^x_i\equiv 0$, $\delta P^{d,s',x}_{CDW}=\delta \Delta^{d,s',x}_{PDW}\equiv 0$

\item Coexistent pair-density-wave and charge-density-wave phase (PDW+CDW): $\bar{P}^{d,s',s'',x,x'}\neq 0$, $\bar{\Delta}^{d,s',x}\neq 0$, $\delta P^{d,s',s'',x,x'}_{CDW}\neq 0$, $\delta \Delta^{d,s',x}_{PDW}\neq 0$
\end{itemize}


\section{Results and discussion}
First, we analyze the appearance of all the previously defined phases (PM, SC, SC+N, CDW+PDW) for the case of the Hubbard model and subsequently, we discuss the effect of adding the exchange interaction term $\sim J$, what leads to the $t$-$J$-$U$ model. In that formulation the $t$-$J$ model is recovered as $U\rightarrow \infty$ ($U>>t$) limit.

\begin{figure}[h!]
\centering
\includegraphics[width=1.0\textwidth]{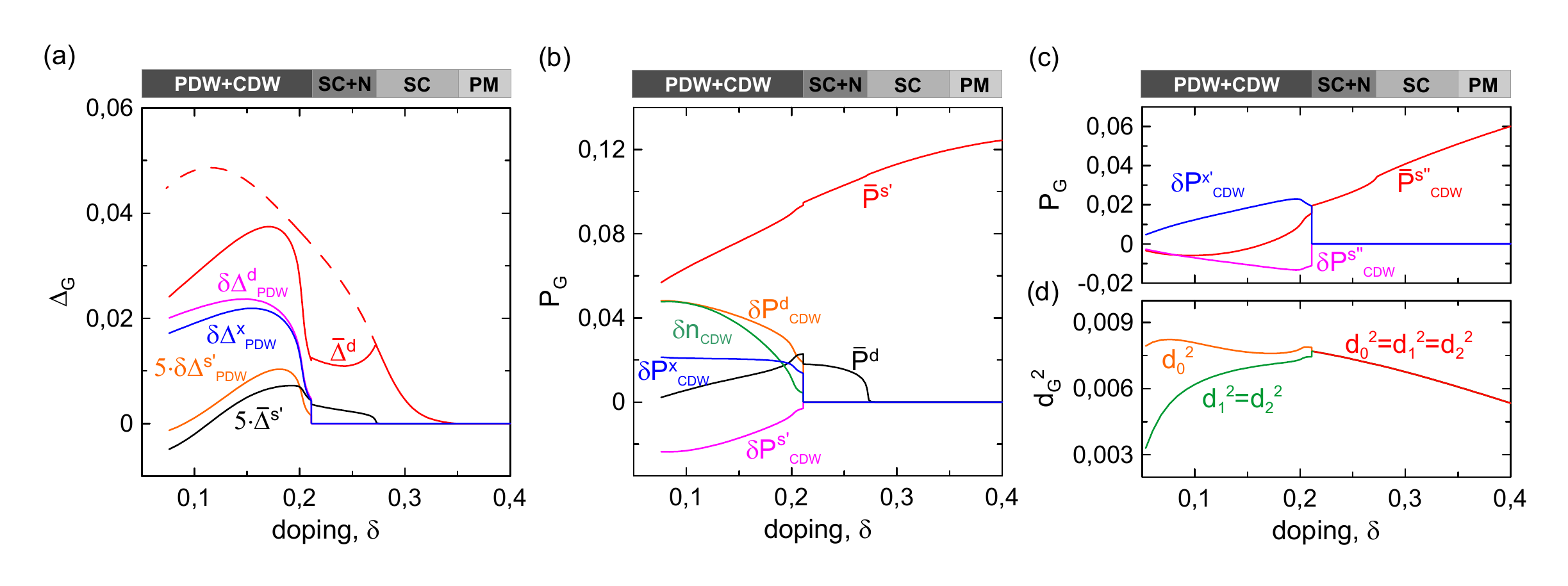}
\caption{Phase diagram for the case of the Hubbard model ($J\equiv 0$) for $U=18$. (a) $d$-$wave$ and $extended$ $s$-$wave$ correlated gaps reference values $\bar{\Delta}^d$, $\bar{\Delta}^{s'}$, as well as the modulation amplitudes $\delta \Delta_{PDW}^d$, $\delta \Delta_{PDW}^{s'}$, and $\delta \Delta_{PDW}^x$ [cf. Eqs. (\ref{eq:mod_delt_ds}) and (\ref{eq:mod_delt_p})], all as functions of doping. Non-zero values of both $\bar{\Delta}^d$ and $\bar{\Delta}^{s'}$ signal the appearance of $C_4$ symmetry breaking while $\delta \Delta_{PDW} \neq 0$ is the evidence of SC gap modulation in real space, what in turn leads to the pair-density-wave state appearance. (b) and (c) Extended $s$-$wave$ and $d$-$wave$ hopping base values $\bar{P}^{s'}$, $\bar{P}^{s''}$ and $\bar{P}^d$, respectively, as well as the modulation amplitudes $\delta P_{CDW}^d$, $\delta P_{CDW}^{s'}$, $\delta P_{CDW}^{s''}$, $\delta P_{CDW}^x$, and $\delta P_{CDW}^{x'}$ [cf. Eqs. (\ref{eq:mod_P_dsp}) and  (\ref{eq:mod_P_p})] all as functions of doping. Non-zero values of both $\bar{P}^d$ and $\bar{P}^{s'}$ also signal the appearance of $C_4$ symmetry breaking while $\delta P_{CDW}\neq 0$ is the evidence of the average hopping modulations in real space, what in turn leads to the bond-centered charge-density-wave state. Additionally, for $\delta n_{CDW}\neq 0$ the site-centered charge-density-wave appears. (d) Double occupancies in the correlated state corresponding to the atomic sites labeled (0), (1), and (2) (cf. Fig. \ref{fig:lattice}).}
\label{fig:ndep_Hub}
\end{figure}

In Fig. \ref{fig:ndep_Hub} we display the correlated superconducting gaps (reference values $\bar{\Delta}$ and amplitudes $\delta \Delta_{PDW}$), correlated hoppings (reference values $\bar{P}$ and amplitudes $\delta P_{CDW}$), amplitude of particle number modulations in real space ($\delta n_{CDW}$), as well as double occupancies, all as a function of hole doping for the case of the Hubbard model with $U=18$ [$J\equiv 0$ in Hamiltonian (\ref{eq:H_start})]. In the left part of the Figure we mark the stability regimes of particular phases. By going from the high-doping side we first encounter the paramagnetic phase (PM) with vanishing superconducting gap, no charge ordering and the $C_4$ symmetry conserved. For the dopings below $\delta\approx 0.35$ a pure $d$-$wave$ superconducting phase (SC) is stable ($\bar{\Delta}^d\neq 0$). After passing the value $\delta\approx 0.28$, the $s$-$wave$ component of the SC gap ($\bar{\Delta}^s$) and the $d$-$wave$ correlated hopping component ($\bar{P}_d$) become non-zero, which signals the appearance of nematicity coexisting with the paired state (SC+N). An extensive analysis of the coexisting nematic-superconducting phase within the Hubbard and the $t$-$J$-$U$ models is provided in Ref. \cite{Zegrodnik2017_nematic}. The transition from SC to SC+N phase is of the second order. One can see that in the SC+N phase the $d$-$wave$ component of the SC gap is reduced with respect to the case of pure $d$-$wave$ superconductivity, which is marked by the red dashed line in Fig. \ref{fig:ndep_Hub} (a). In the nematic phase the $C_4$ symmetry is spontaneously broken and the $(1,0)$ and $(0,1)$ directions are no longer equivalent in the electronic wave function, even though the underlying crystal lattice has no distortion (is still square). Moreover, in such a phase the translational symmetry is conserved which means that no charge or SC gap modulation appears as yet ($\delta P_{CDW}=\delta n_{CDW}=\delta \Delta_{PDW}=0$). 

The SC+N phase can be understood as a precursor of the PDW+CDW phase, (for which the $C_4$ symmetry is also broken), which appears below the doping $\delta\approx 0.2$, at which a first order transition takes place. In the PDW+CDW phase the SC gap, the average number of electrons, as well as the hopping averages are all modulated along the $x$ axis with the modulation vector $\mathbf{Q}=(1/3,0)2\pi$. In this phase the pair-density wave coexists with both the site-centered and the bond-centered charge orderings and all the modulation amplitudes have non-zero values [$\delta P_{CDW}\neq 0$, $\delta n_{CDW}\neq 0$, $\delta \Delta_{PDW}\neq 0$, cf. Fig. \ref{fig:ndep_Hub} (a), (b), and (c)]. For the sake of completeness, in Fig. \ref{fig:ndep_Hub} (d) we plot the double occupancies corresponding to the three sublattices, which compose the pattern along the $x$ direction. 

\begin{figure}[h!]
\centering
\includegraphics[width=1.0\textwidth]{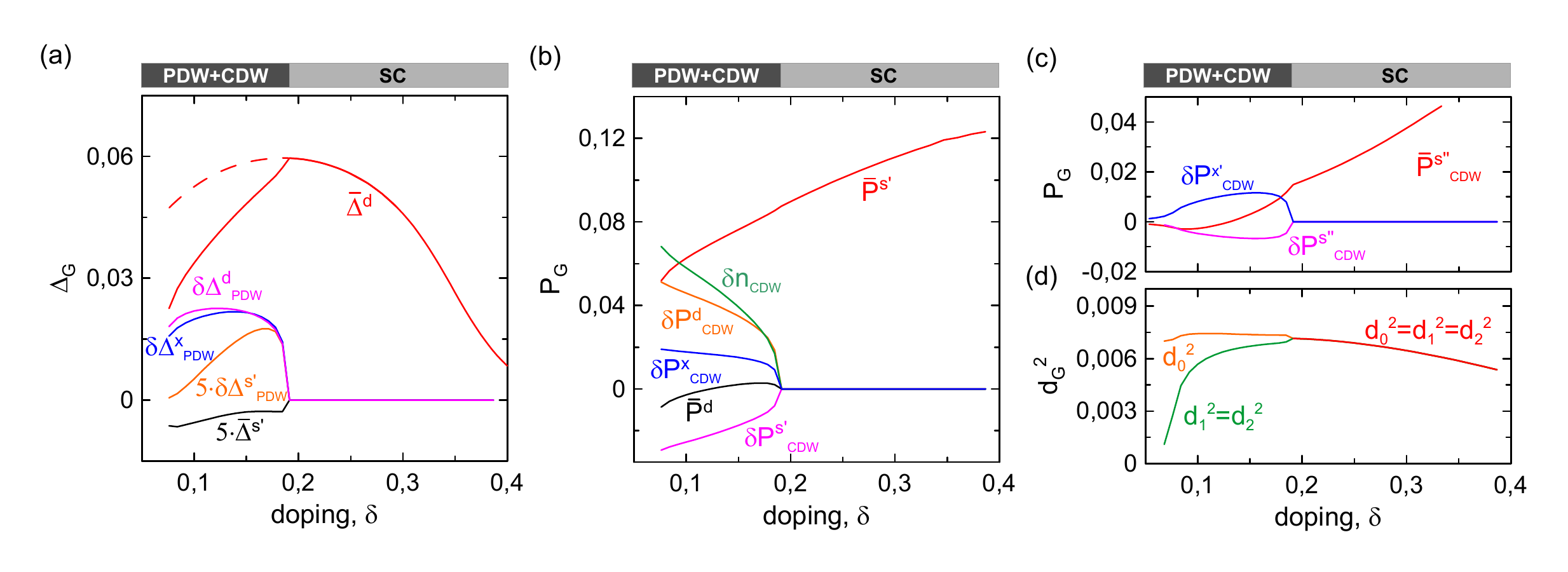}
\caption{Phase diagram for the case of the $t$-$J$-$U$ model with $J=0.3$ and $U=18$. (a) $d$-$wave$ and $extended$ $s$-$wave$ correlated gaps reference values $\bar{\Delta}^d$, $\bar{\Delta}^{s'}$, as well as the modulation amplitudes $\delta\Delta_{PDW}^d$, $\delta\Delta_{PDW}^{s'}$, and $\delta\Delta_{PDW}^x$ [cf. Eqs. (\ref{eq:mod_delt_ds}) and (\ref{eq:mod_delt_p})], all as functions of doping. $\delta \Delta_{PDW} \neq 0$ is the evidence of SC gap modulation in real space, what in turn leads to the pair-density-wave state appearance. (b) and (c) $Extended$ $s$-$wave$ and $d$-$wave$ hopping reference values $\bar{P}^{s'}$, $\bar{P}^{s''}$ and $\bar{P}^d$, respectively, as well as the modulation amplitudes $\delta P_{CDW}^d$, $\delta P_{CDW}^{s'}$, $\delta P_{CDW}^{s''}$, $\delta P_{CDW}^x$, and $\delta P_{CDW}^{x'}$ [cf. Eqs. (\ref{eq:mod_P_dsp}) and  (\ref{eq:mod_P_p})] all as functions of doping. For dopings for which $\delta P_{CDW}\neq 0$ ($\delta n_{CDW}\neq 0$) the bond-centered (site-centered) charge-density-wave state sets in. (d) Double occupancies in the correlated state corresponding to the atomic sites labeled (0), (1), and (2) (cf. Fig. \ref{fig:lattice}).}
\label{fig:ndep_tJU}
\end{figure}

The results for the case of the $t$-$J$-$U$ model are presented in Fig. \ref{fig:ndep_tJU}. As it has already been reported in Ref. \cite{Zegrodnik2017_nematic} the exchange term has a negative effect on the nematicty while positive on the superconductivity. The same can be seen here: for the $J$ value representative for the cuprates ($J\approx 0.3$) the nematic phase is completely suppressed while the stability of the superconducting phase is extended to higher doping values with respect to the Hubbard model case (cf. Fig. \ref{fig:ndep_Hub}). As a result, the pure $d$-$wave$ SC phase is stable down to the optimal doping value $\delta\approx 0.18$ at which a second order phase transition appears to the PDW+CDW state without the appearance of the precusor nematic phase in between the SC and PDW+CDW stability regimes. The red dashed line in Fig. \ref{fig:ndep_tJU} (a) marks the continuation of the $d$-$wave$ SC gap for the case when the PDW+CDW stability is not included in the calculations. In such a case the value $\delta\approx 0.18$ corresponds to the maximal correlated gap and hence the maximal critical temperature. The stability of the PDW+CDW phase in the underdoped regime obtained in our calculations reflects the experimental findings for the cuprates, where the charge ordered phase is observed in a similar doping range \cite{Comin2014-Science,Tabis2014,Blanco2014}. The SC gap parameters modulation amplitudes in the PDW+CDW phase are shown in Fig. \ref{fig:ndep_tJU} (a) and have a dome-like shape contained in the underdoped regime with a maximum value at $\delta\approx 0.13$. On the other hand, the amplitudes of electron hopping modulation ($\delta P_{CDW}$) and electron concentration modulation ($\delta n_{CDW}$) are increasing with the decreasing doping which is in correspondence with the measured doping dependence of the CDW critical temperature \cite{Chang2012,Tabis2014}. One should note that the parameter set corresponding to Fig. \ref{fig:ndep_tJU} is close to the one for which good agreement between theory and experiment has been obtained with respect to selected universal properties of the pure superconducting phase in the cuprates \cite{Spalek2017}.

As can be seen from Figs. \ref{fig:ndep_Hub} and \ref{fig:ndep_tJU} both the $s$-$wave$ ($\delta n_{CDW}$), $extended$ $s$-$wave$ ($\delta P^{s'}_{CDW}$, $\delta\Delta^{s'}_{PDW}$) and $d$-$wave$ ($\delta P^{d}_{CDW}$, $\delta\Delta^{d}_{PDW}$) contributions to the pairing and hopping modulations have non-zero values in the obtained CDW+PDW phase. According, to the experimental analysis in the large group of the copper-based compounds the dominant $d$-$wave$ form factor is believed to appear. Our calculations show that the $d$-$wave$ form factor amplitudes are significantly larger than the $extended$ $s$-$wave$ correspodants. However, the site-centered $s$-$wave$ contribution to the CDW ($\delta n_{CDW}$) is still quite significant within our analysis [cf. Figs. \ref{fig:ndep_Hub} (a) and (b), as well as Figs. \ref{fig:ndep_tJU} (a) and (b)]. Also, we obtain the $\delta \Delta^x_{PDW}$ and $\delta P^x_{CDW}$ form-factors, which reflect the inequivalence between the hopping/pairing to the left- and right-hand side neighbors of the atomic sites (1) and (2). Such a description is necessary in order to obtain the considered periodicity of the hopping and pairing averages.

\section{Conclusions}

The present paper concludes our constuction of a fairly complete phase diagram for the high-T$_C$ cuprate superconductors within a single-band DE-GWF scheme \cite{Spalek2017, Zegrodnik2017_1, Zegrodnik2017_2, Zegrodnik2017_nematic}. Namley, we have analyzed the coexistence of the pair- and charge-density-wave phases within the Hubbard and $t$-$J$-$U$ models by using the DE-GWF method. The calculations have been carried out for the fixed modulation vector $\mathbf{Q}=(2\pi/3,0)$, which specifies both the pair- and the charge-density-wave periodicities (as it was reported in Ref. \cite{Hamidian2016}) and is close to that measured experimentally.

Results corresponding to the $t$-$J$-$U$ model confirm the stability of a pure $d$-$wave$ superconducting phase down to the hole doping $\delta\approx 0.18$, which corresponds to the maximal correlated gap, and therefore is identified as the optimal doping. Below that value the coexistent PDW+CDW phase sets in for which both the $C_4$ and translational symmetries are spontaneously broken. The $d$-$wave$ SC gap parameter is reduced in the PDW+CDW stability regime and the PDW pairing modulation amplitudes form a dome-like shape confined within the underdoped regime. On the other hand, the CDW hopping and electron concentration modulation amplitudes increase with the decreasing doping, what is reminiscent of the $T_{CDW}$ doping dependence determined by the X-ray diffraction experiments \cite{Chang2012}. Also, the fact that the modulated phase appears in the underdoped regime and that the $d$-$wave$ symmetry modulation form-factor is significantly larger than that of the $extended$ $s$-$wave$, agrees with the experimental observations. Nevertheless, our approach leads also to a significant contribution of the site-centered CDW ordering which means that the zero-gap state in the nodal direction is lost. 

For the case of the Hubbard model a narrow stability range of the coexistent superconducting-nematic phase appears in between the pure superconducting and the PDW+CDW phases. In the SC+N phase the rotational symmetry is broken, however, the translational symmetry is conserved. Such a phase can be considered as a precursor state for the formation of the CDW+PDW phase when decreasing the doping. The absence of the SC+N state for the case of the $t$-$J$-$U$ model is due to the negative influence of the exchange term $\sim J$ on the nematic phase, which  was reported recently \cite{Zegrodnik2017_nematic,Kaczmarczyk2016}.

Note that in related analysis the intersite Coulomb interaction was included in order to induce the charge-ordered state \cite{Allais2014,Abram2017,Amaricci2010,Terletska2017,Kapcia2017}. Within that approach the appearance of CDW can be understood in a straightforward manner. Namely, in the simplest case of site-centered checkerboard pattern, the role non-local electron repulsion ($\sim V$) is minimal at the cost of the increasing local interaction energy ($\sim U$). In such a case the charge ordering appears after reaching the critical $V$ value \cite{Abram2017,Amaricci2010,Terletska2017,Kapcia2017}. For the case of more sophisticated charge orderings, including also bond order, the situation is not so intuitivly clear, especially when an additional pair-density-wave modulation comes into play. From our present analysis it follows that the intrasite Coulomb interaction is sufficient to induce both the pure $d$-$wave$ SC as well the charge/pair modulated states; hence, the $V$ term is absent in our approach. However, the stability of the mentioned phases could not be reproduced within our zeroth-order diagrammatic expansion (\ref{eq:expansion}), which is equivalent to RMFT. This means that the higher order terms of the DE-GWF are instrumental for the spontaneous symmetry breaking, here expressed in terms of the formation of SC, PDW/CDW states. The appearance of the bond-ordered state coexisting with $d$-$wave$ SC induced purely by local Coulomb repulsion has also been reported recently in Ref. \cite{Faye2017} but the results differ form ours.

The principal conclusion coming from the two models (Hubbard and $t$-$J$-$U$) studied here is that the $t$-$J$-$U$ model provides results which are in closer correspondence to the experimental data for the CDW+PDW modulated state of the cuprates. In our very recent paper we have shown that this model also leads to a good quantitative agreement between theory and experiment for the selected universal characteristics of the pure $d$-$wave$ superconducting state \cite{Spalek2017}. The parameter set taken in the earlier analysis \cite{Spalek2017} is close to that considered here. Nevertheless, the proper balance between the symmetry form-factors of the CDW+PDW phase is still lacking within the single-band DE-GWF description. The dominant $d$-$wave$ bond-ordered phase, which is believed to appear in many copper-based compounds, can be ascribed to a modulating charge located on the oxygen $2p$ orbitals of the Cu-O plane \cite{Comin2015,Achkar2016}. To incorporate explicitly such a scenario one has to consider a more realistic 3-band $d$-$p$ model. However, the application of the DE-GWF method to such a case introduces a degree of complexity difficult to handle at present time. We should see progress along this line in the near future, as only then we can reliably estimate the limits of applicability of the one-band effective models.

\section{Acknowledgement}

The discussions with Maciek Fidrysiak and Marcin Abram are gratefully acknowledged. MZ acknowledges the finantial support through the Grant SONATA, No. 2016/21/D/ST3/00979 from the National Science Centre (NCN), Poland. JS acknowledges the financial support through the Grant MAESTRO, No. DEC-2012/04/A/ST3/00342 from the National Science Centre (NCN) of Poland.

\section*{References}

\end{document}